# $sin[n\Delta t\, sin(n\Delta t_1)]$ as a source of unpredictable dynamics

Stefano Morosetti

Department of Chemistry, University of Rome 'La Sapienza', P.le A. Moro 5, 00185

Rome, Italy

e-mail: stefano.morosetti@uniroma1.it, phone: +390649913730, fax: +3906490631

**Abstract**

We investigate the ability of the function $sin[n\Delta t\, sin(n\Delta t_1)]$, where $n$ is an integer and growing number, to produce unpredictable sequences of numbers. Classical mathematical tools for distinguishing periodic from chaotic or random behaviour, such as sensitivity to the initial conditions, Fourier analysis, and autocorrelation are used. Moreover, the function $cos^{-1}\{sin[n\Delta t\, sin(n\Delta t_1)]\}/\pi$ is introduced to have an uniform density of numbers in the interval [0,1], so it can be submitted to a battery of widely used tests for random number generators. All these tools show that a proper choice of $\Delta t$ and $\Delta t_1$, can produce a sequence of numbers behaving as unpredictable dynamics.

**Keywords**: Chaotic systems; Random systems; Random number generator

**1. Introduction**

Recently a class of functions which produce unpredictable dynamics has been shown [2-5,8]. These functions can be written in the form

$$X_n = h[f(n)] \qquad (1)$$

where the argument function $f(n)$ grows exponentially and the function $h(y)$ should be finite, and noninvertible. In other words $h(y) = \alpha$ with $\alpha$ constant, possesses several solution for $y$. An example is

$$X_n = [sin\, \theta_z{}^n]^2 \qquad (2)$$

where $n$ is an integer and growing number.

The function $f(n)$ does not have to grow exponentially all the time. It is sufficient for $f(n)$ to possess repeating intervals with finite exponential behaviour. The exponential growing imposes the necessity of often rescale the $f(n)$ function in order to implement it on a computer, where the storable maximum number is finite in size if a predefined number of bits is used for storage.

We investigate the ability of the function

$$sin[n\Delta t \, sin(n\Delta t_1)] \qquad (3)$$

where $n$ is an integer and growing number, to produce unpredictable sequences of numbers and to have chaotic behaviour. In this case the argument grows linearly and not exponentially as in Eq. (2). This allows an easier tractability of the argument.

## 2. Testing the $sin[n\Delta t \, sin(n\Delta t_1)]$ function

The appearance of the function is shown in Fig. 1, where a continuous $t$ is used instead of $n\Delta t$ and $n\Delta t_1$.

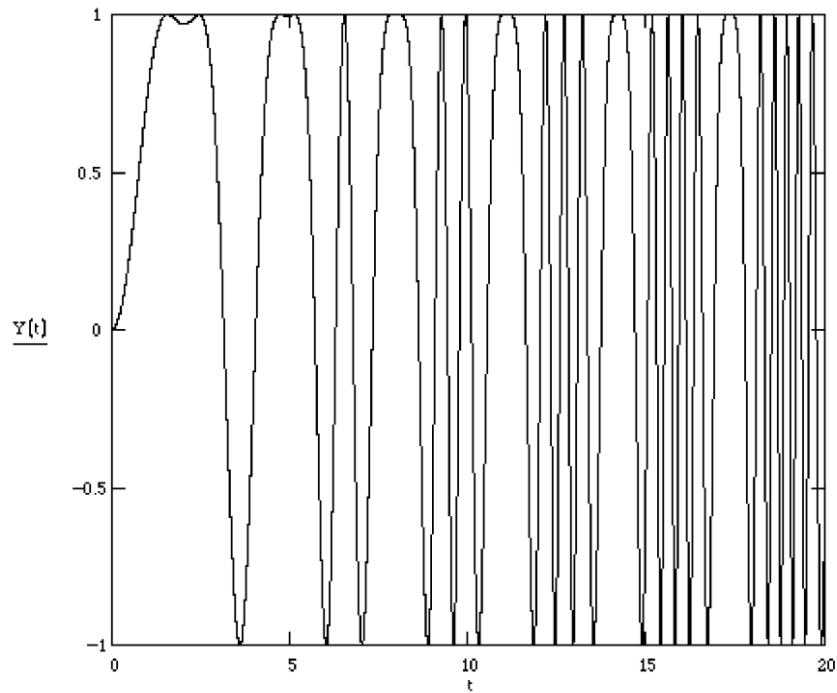

Fig. 1. The frequency of the function $Y(t) = sin[t \, sin(t)]$ grows with $t$.

In calculating the $sin(t)$, the argument $t$ is subject to an operation of module in respect to $2\pi$, and this guarantees that a growing argument as $n\Delta t$ will be projected into a finite interval.

We distinguish in the following the $sin$ operator as internal (inside the square brackets in function (3)) or external (outside the square brackets).

The growing of the $n\Delta t$ term can be roughly seen as a growing frequency of the $sin$ function. In fact, if $t = n\Delta t$ is so high that $t \cong t + 2\pi$, while the inside $sin$ will range from -1 to +1 during a cycle, the argument of the outside $sin$ will range from $-t$ to $+t$, and as a consequence the number of cycles of the outside sin will be $2t/2\pi$, and its approximate frequency will be $t/\pi$.

In the following, classical mathematical tools for distinguishing periodic from chaotic or random behaviour, such as Fourier analysis, sensitivity to the initial conditions, and autocorrelation will be used. Moreover, a modification of the function will be introduced to have uniform density of numbers in the interval [0,1], so it will be submitted to a battery of widely used tests for random number generators.

### 2.1 Fourier analysis

A mathematical analysis tool for distinguishing periodic from chaotic or random behaviour is Fourier analysis.

The function (3) is equivalent to a Fourier sine series with Bessel functions as coefficients [1]:

$$sin[n\Delta t \, sin(n\Delta t_1)] = 2 \sum_{k=0}^{\infty} \{[J_{2k+1}(n\Delta t)] * sin[(2k+1)n\Delta t_1]\} \qquad (4)$$

and from Eq. (4) it can be expected that function (3) has infinite Fourier components, except for particular choices of $\Delta t$ and $\Delta t_1$.

A sequence of numbers can be generated by assigning to $n$ the natural series of integer numbers.

We explore under what conditions the function (3) behaves periodically and then produces a periodic sequence of numbers. To obtain the same value at different times, the following condition must be satisfied:

$$\sin[n_1 \Delta t \sin(n_1 \Delta t_1)] = \sin[n_2 \Delta t \sin(n_2 \Delta t_1)]$$

hence:

$$n_2 \Delta t \sin(n_2 \Delta t_1) = 2k\pi + n_1 \Delta t \sin(n_1 \Delta t_1) \qquad (5)$$

or

$$n_2 \Delta t \sin(n_2 \Delta t_1) = (2k+1)\pi - n_1 \Delta t \sin(n_1 \Delta t_1) \qquad (6)$$

where $k$ is an integer.

We set the following equalities to switch from discrete to continuous values:

$$n_1 \Delta t = t \qquad n_2 \Delta t = (n_1 + n_3)\Delta t = (t + \tau) \qquad \Delta t_1 = \alpha \Delta t$$

$$n_1 \Delta t_1 = \alpha t \qquad n_2 \Delta t_1 = (n_1 + n_3)\alpha \Delta t = \alpha(t + \tau)$$

where $t$ is the variable and both $\tau$ and $\alpha$ are costants.

Replacing in 5) we obtain:

$$(t + \tau) \sin[\alpha(t + \tau)] = (2k)\pi + t \sin(\alpha t)$$

$$\downarrow$$

$$t \sin[\alpha(t + \tau)] + \tau \sin[\alpha(t + \tau)] - t \sin(\alpha t) = (2k)\pi$$

$$\downarrow$$

$$t\{\sin[\alpha(t + \tau)] - \sin(\alpha t)\} + \tau \sin[\alpha(t + \tau)] = (2k)\pi$$

Using the Prosthaphaeresis formula:

$$2t \cos\left(\frac{\alpha t + \alpha\tau + \alpha t}{2}\right) \sin\left(\frac{\alpha t + \alpha\tau - \alpha t}{2}\right) + \tau \sin[\alpha(t + \tau)] = (2k)\pi$$

$$\downarrow$$

$$2t \cos\left[\alpha\left(t + \frac{\tau}{2}\right)\right] \sin\left(\frac{\alpha\tau}{2}\right) + \tau \sin[\alpha(t + \tau)] = (2k)\pi$$

The term $\tau \sin[\alpha(t + \tau)]$ will range between the constant values $-\tau$ and $+\tau$, while $2t \cos\left[\alpha\left(t + \frac{\tau}{2}\right)\right] \sin\left(\frac{\alpha\tau}{2}\right)$ will range between $-2t$ and $+2t$, where $t$ is the increasing variable. This last term oscillates, but with maximum and minimum that tend respectively to $+\infty$ and $-\infty$. $(2k)\pi$ being a finite term, the equality can be satisfied only by deleting the $2t \cos\left[\alpha\left(t + \frac{\tau}{2}\right)\right] \sin\left(\frac{\alpha\tau}{2}\right)$ term, that is, by setting $\tau$ or $\alpha$ to 0 and $k$ to 0. For $\alpha = 0$ the functions are 0. If $\tau = 0$, then $n_1 \Delta t = n_2 \Delta t$ and $n_1 \Delta t_1 = n_2 \Delta t_1$, therefore the two functions are identical.

Considering the relation 6), the same procedure leads to the following result:

$$2t \sin\left[\alpha\left(t + \frac{\tau}{2}\right)\right] \cos\left(\frac{\alpha\tau}{2}\right) + \tau \sin[\alpha(t + \tau)] = (2k + 1)\pi$$

The same condition of a finite function can be satisfied by setting $\alpha = 0$, or $\alpha\tau = (2k_1 + 1)\pi$, with $k_1$ integer. In either case the equality cannot be satisfied: in the $\alpha = 0$ case, the members on the left are 0, but $(2k + 1)\pi$ cannot be 0 for any integer value of $k$; in the $\alpha\tau = (2k_1 + 1)\pi$ case, the first member on the left is 0, the second member becomes $\tau \sin[\alpha t + (2k_1 + 1)\pi]$, which is not constant, so it cannot be equal to $(2k + 1)\pi$. In conclusion the relation 6) cannot be satisfied.

We can conclude the function 3) is always aperiodic.

From the properties of the function, i.e. it is aperiodic, with growing frequency and noninvertible, it follows that the sequence $X_n$, generated by assigning to $n$ in the Eq. (3) the natural series of integer numbers, is in general unpredictable, in the sense that given any string of $m + 1$ values $X_0, X_1, X_2, \ldots X_m$, the next value $X_{m+1}$ is not determined by knowing the previous values (when the integers $i$ used to obtain the $X_i$ values are unknown).

## 2.2 First-return map

The first-return map is another tool for visualising the chaotic behaviour of the number sequence.

We produce the values $X_0, X_1, X_2, \ldots$ with the function (3) (see Fig. 2) and the first-return map can be constructed from this sequence (see Fig. 3).

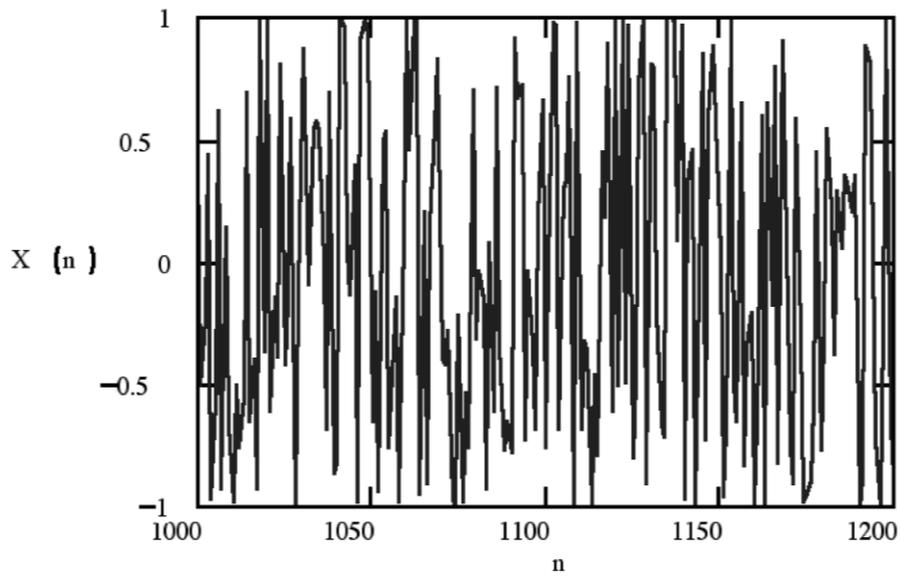

Fig. 2 . Numbers sequence obtained by the function (3), where $\Delta t = 3$, $\Delta t_1 = 1.01 \Delta t$, $n$ ranging from 1000 to 4000. A line connecting the numbers, shows a very complex behaviour.

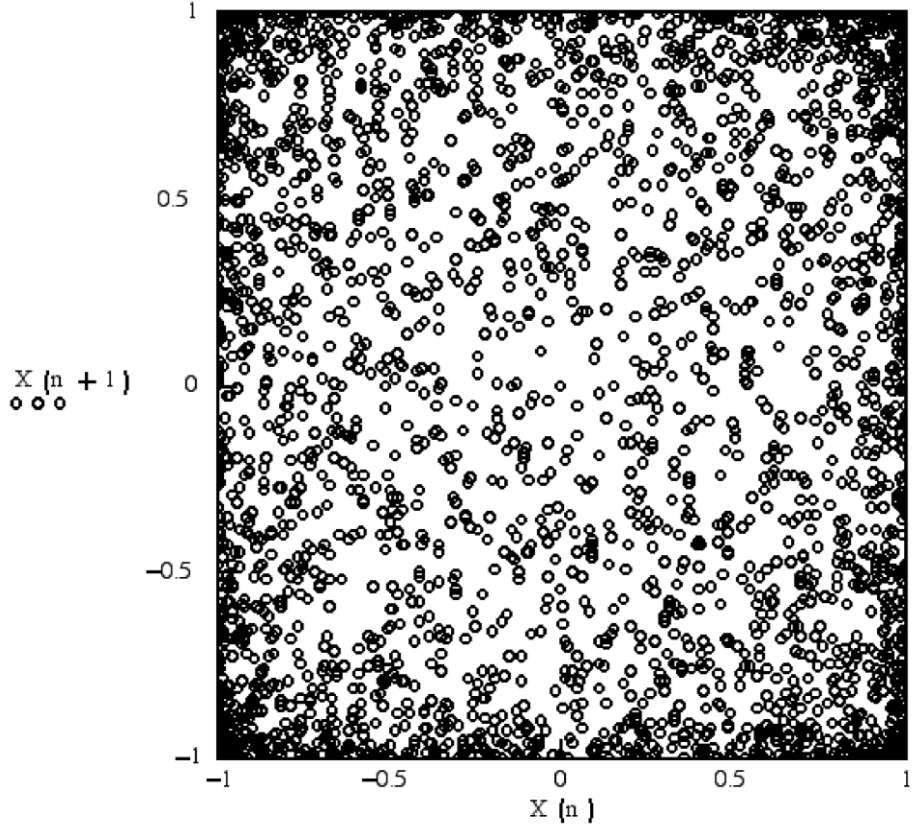

Fig. 3. First return map produced by the function (3). $\Delta t = 3$, $\Delta t_1 = 1.01\Delta t$, $n$ ranging from 1000 to 4000. Note that the dynamics is structureless.

To produce the chaotic sequences we use a $\Delta t$ which is in an irrational quotient with $\pi$, taking into account the above considerations. Moreover the first number assigned to $n$ is high (>1000), and $\Delta t > \pi/10$. These last technical choices follow from the time ($n\Delta t$) necessary to have the maximum divergence (that is 2 in Eq. (3)) in 2 near trajectories, which is calculated later.

## 2.3 Probability density

The numbers produced by function (3) are not distributed uniformly, as can be seen also in Fig. 3. The probability density behaves as $P(X) \cong 1/(1 - X^2)^{1/2}$ (See Fig. 4).

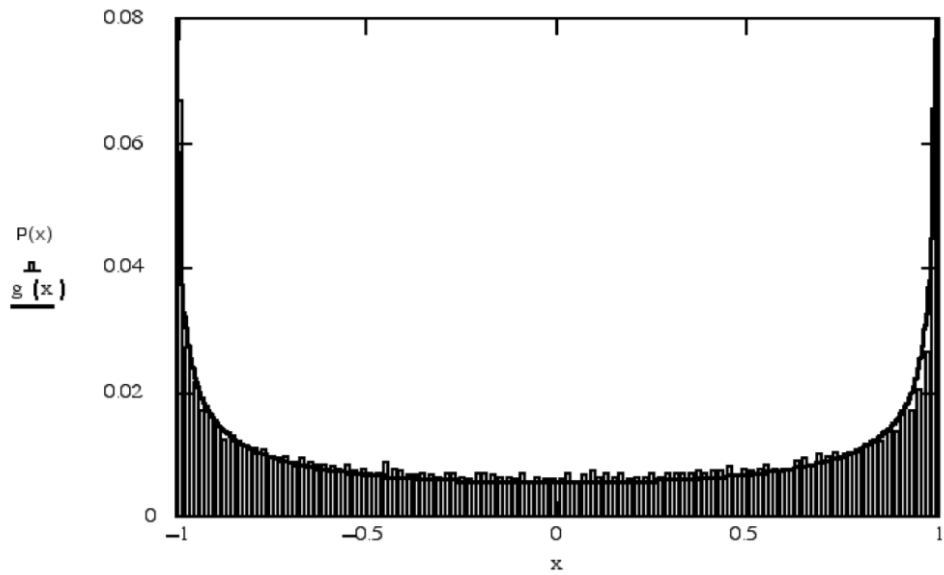

Fig. 4. $P(X)$ is the density distribution for histograms of the values obtained by the function (3). $\Delta t = 3$, $\Delta t_1 = 1.01\Delta t$, $n$ ranging from 0 to 33000. The $x$ range is divided into 100 intervals. $g(x)$ is the better interpolating function of the type $P(X) = 1/(1 - X^2)^{1/2}$.

If we need uniformly distributed numbers, we should make the following transformation:

$$Y_n = acos(X_n)/\pi \qquad (7)$$

obtaining $P[Y(X)] = cost$ (see Figs. 5-6).

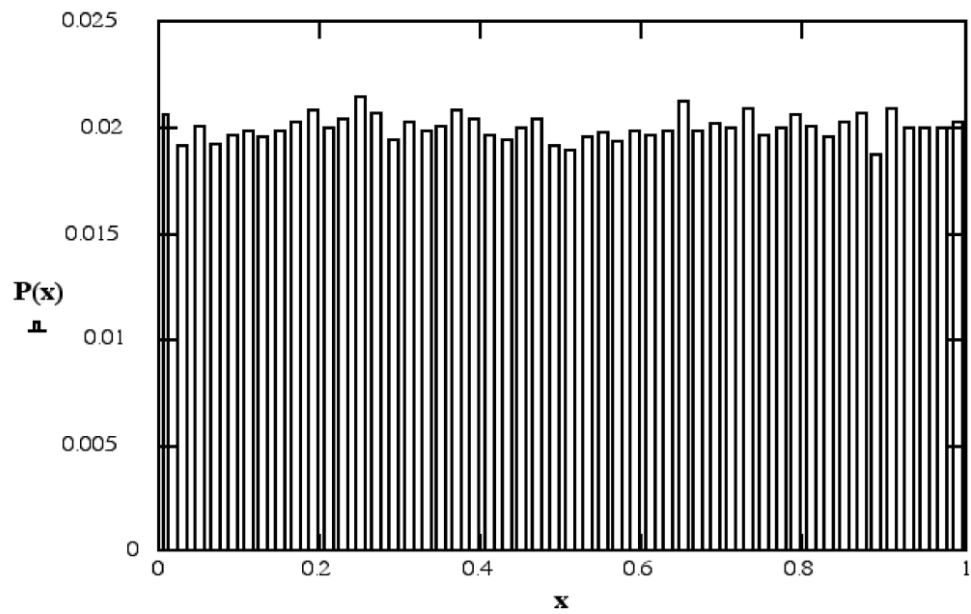

Fig. 5. $P(X)$ is the density distribution for histograms of the values obtained by the function (7). $\Delta t = 3$, $\Delta t_1 = 1.01\Delta t$, $n$ ranging from 0 to 33000. The $x$ range is divided into 50 intervals.

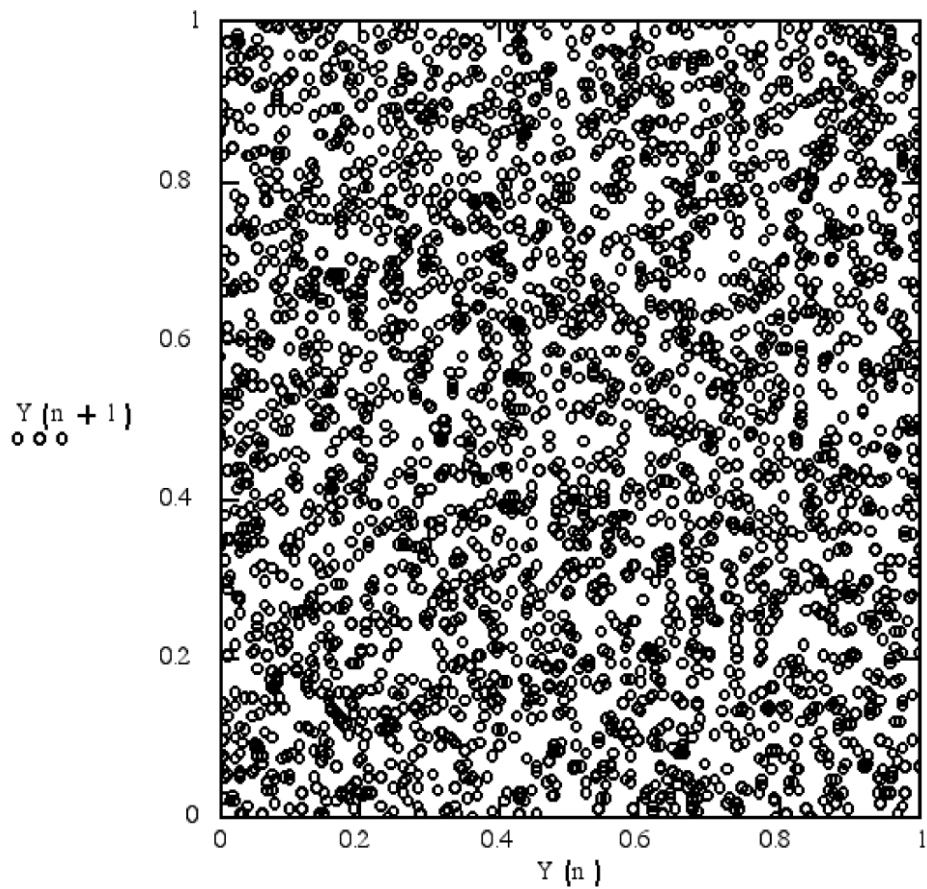

Fig. 6. As Fig. 3 but the use of function (7).

## 2.4 The sensitivity to the initial conditions

To confirm the chaotic behaviour of the sequences of numbers obtained through the Eqs. (3) and (7), we investigate the sensitivity to the initial conditions.

In Fig. 7 a difference in Eq. (3) of a factor of only $1*10^{-6}$ in $\Delta t$ produces very different first return maps.

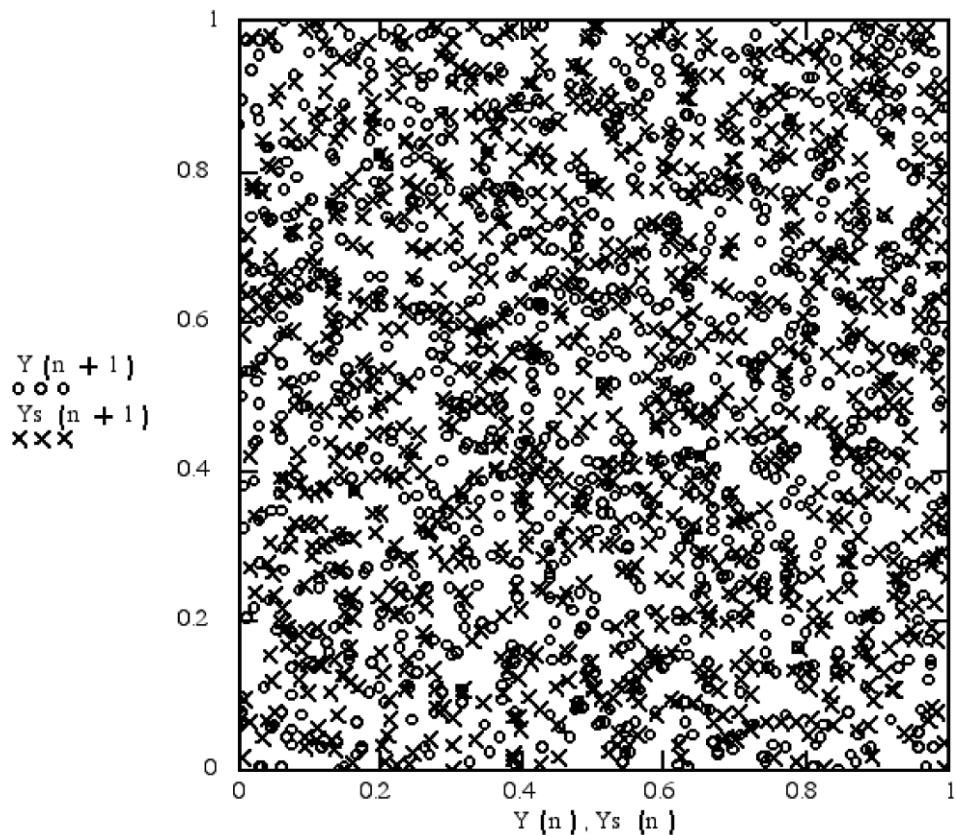

Fig. 7 . First return map produced by the function (3). $n$ ranging from 1000 to 2000. Circle: $\Delta t = 3$, $\Delta t_1 = 1.01\Delta t$. Cross: $\Delta t = 3 * 1.000001$, $\Delta t_1 = 1.01\Delta t$.

In Fig. 8 we report the mean distance for 5 pairs of nearby trajectories obtained through the Eq. (7), where all the trajectories are obtained assigning to $n$ the natural integer number sequence. For all the functions $\Delta t = 3$, and $\Delta t_1 = 1.01\Delta t$. The different trajectories are obtained adding different values to the arguments of the sines. In each pair, the near trajectory is obtained adding a further value of $10^{-6}$ to the previous one.

The result allows a rough estimate of the higher Lyapunov exponent, which results to be approximately 1.

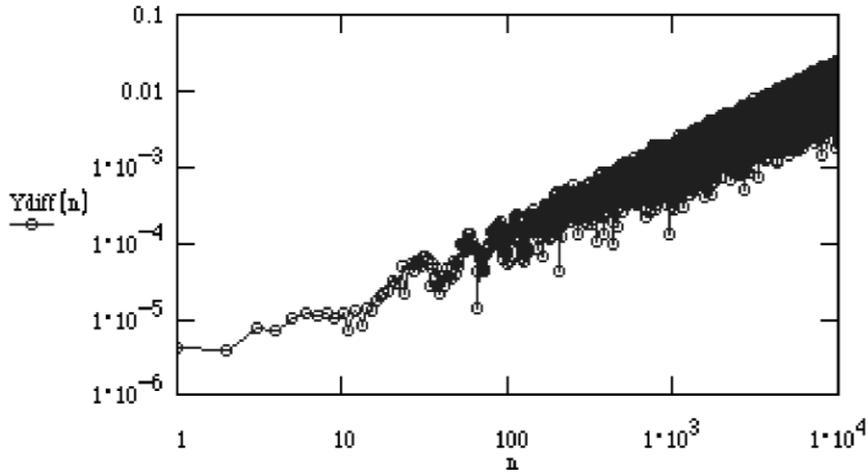

Fig. 8. $n$ represents the natural sequence of the integer numbers. 5 pairs of nearby trajectories obtained as sequences of the values calculated from function (7) are considered; for all the functions $\Delta t = 3$, and $\Delta t_1 = 1.01\Delta t$; the different trajectories are obtained adding different values to the arguments of the sines, that is: 1., $1/\pi$, $\pi$, $e$, $e + \pi$, respectively. The nearby trajectories are obtained in all cases adding $10^{-6}$ to the previous values. $Ydiff(n)$ is the mean of the differences calculated for each pair of nearby trajectories.

## 2.5 Power spectrum and autocorrelation

In Fig. 9 the Power Spectrum obtained by the Discrete Fourier Transform of a sequence of numbers generated by the function $Z(n) = 2Y(n) - 1$, with $\Delta t = 3$, $\Delta t_1 = 1.01\Delta t$, $n$ ranging from 1 000 to 9 000, where $Y(n)$ is the function (7), is shown. The function $Z(n)$ is chosen to generate numbers in the interval [-1,+1]. No predominant frequency is present.

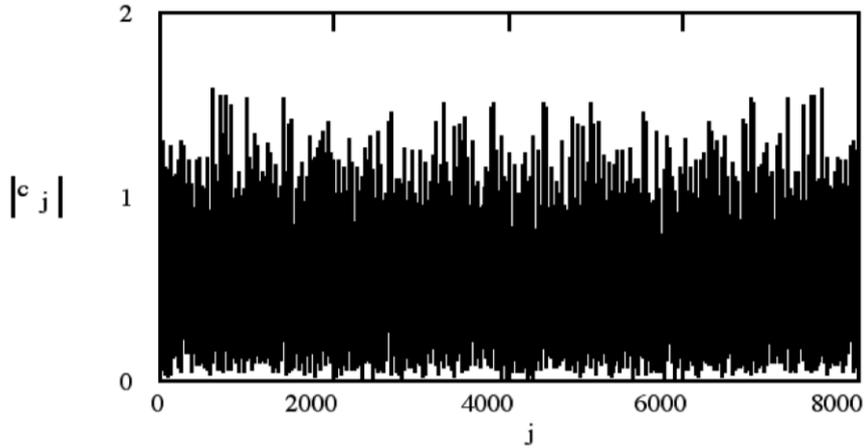

Fig. 9. Power spectrum obtained by the Discrete Fourier Transform of a sequence of numbers generated by the function $Z(n) = 2Y(n) - 1$, where $Y(n)$ is the function (7), with $\Delta t = 3$, $\Delta t_1 = 1.01\Delta t$, $n$ ranging from 1000 to 9000.

The autocorrelation function of $Z(n)$ is defined in Eq. (8) when $f(u)$ is a real function as in our case. $h(x)$ is shown in Fig. 10. Only the zero shift value has a relatively high value, as expected.

$$h(x) = \int_{-\infty}^{+\infty} f(u)f(u+x)du \qquad (8)$$

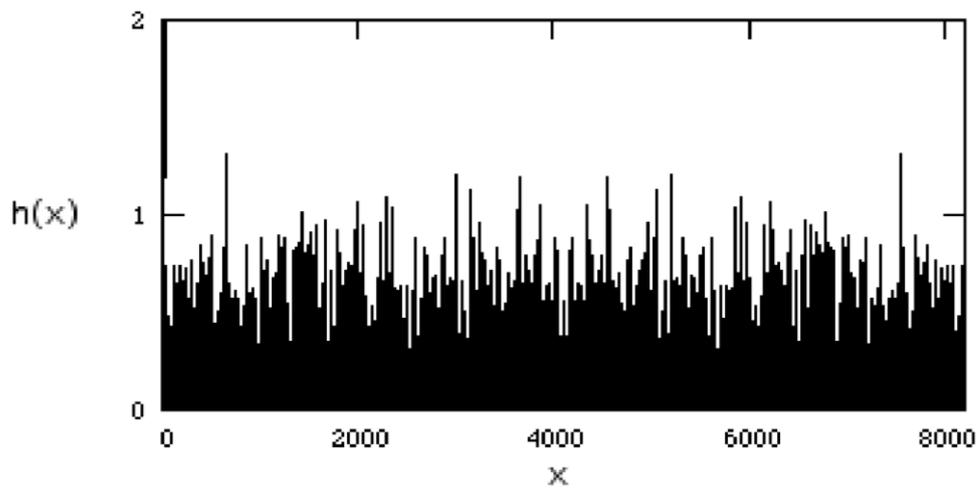

Fig. 10. Autocorrelation function of $Z(n) = 2Y(n) - 1$, where $Y(n)$ is the function (7), with $\Delta t = 3$, $\Delta t_1 = 1.01\Delta t$, $n$ ranging from 1000 to 9000.

## 2.6 Random walk

Recently, simulations of random walks have become a test for pseudo random number generators [7]. We have produced random walks using the numbers generated by the function (7), following the procedure adopted by Nogués [7] for a four-direction movement. The mean distance behaviour possesses the correct property $\langle d^2 \rangle \sim N$ (see Fig.11).

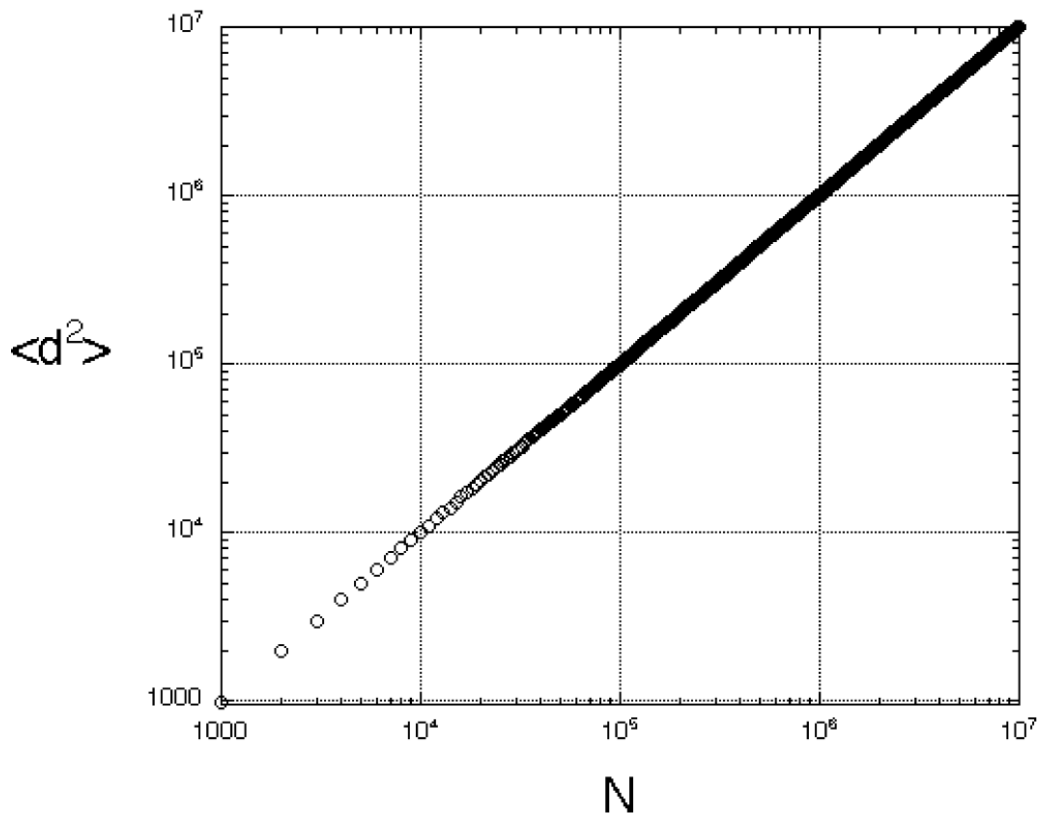

Fig. 11 . Mean distance, averaged over 10 000 different random walks, vs. number of steps. The random walks are generated with the sequence of numbers obtained with the function (7). The random walks are different because $\Delta t$ is random chosen between 1.0 and 3.0. $\Delta t_1 = 1.01 \Delta t$. $n$ starts with the value 1 000.

## 2.7 Tests for random number generators

A library of tests for random number generators is available at the address http://www.csis.hku.uk/cisc/download/idetect/ by the Center for Information Security and Cryptography (CISC) [6]. We have produced sequences of pseudo random numbers using the function (7) multiplied by $2^{32}$-1 to obtain 32 digits numbers, as required for the tests, and with different combinations of $\Delta t$ and $\Delta t_1$. The battery of tests was passed in all cases.

## 3. Results

Using classical mathematical tools for distinguishing periodic from chaotic or random behaviour, as sensitivity to the initial conditions, Fourier analysis, autocorrelation and a battery of widely used tests for random number generators, we have shown that the functions (3), and (7), with a proper choice of $\Delta t$ and $\Delta t_1$, can produce a sequence of numbers behaving as an unpredictable dynamics.